\documentclass[11pt,tightenlines,eqsecnum,floats,aps,amsmath,amssymb,nofootinbib,prd,shownopacs,floatfix]{revtex4}

\usepackage{amsmath,amssymb,amsfonts,dcolumn,color,graphicx,graphics,latexsym}
\usepackage[mathscr]{eucal}
\usepackage[latin1]{inputenc}
\usepackage{enumerate}
\newcommand{\be}{\begin{equation}}
\newcommand{\ee}{\end{equation}}
\newcommand{\ba}{\begin{eqnarray}}
\newcommand{\ea}{\end{eqnarray}}

\newcommand{\lp}{l_{\mathrm{Pl}}}
\newcommand{\f}{\frac}

\newcommand{\bmult}{\nopagebreak[3]\begin{multline}}
\newcommand{\emult}{\end{multline}}

\def\d{{\rm d}}
\def\p{\partial}

\begin{document}

\title{Kantowski-Sachs spacetime in loop quantum cosmology: bounds on expansion and shear scalars and the viability of quantization
prescriptions}
\author{Anton Joe}
\email{ajoe3@lsu.edu}
\author{Parampreet Singh}
\email{psingh@phys.lsu.edu}
\affiliation{ Department of Physics and Astronomy,\\
Louisiana State University, Baton Rouge, LA 70803, U.S.A.}

\begin{abstract}
Using effective dynamics, we investigate the behavior of expansion and shear
scalars in different proposed quantizations of the Kantowski-Sachs spacetime with matter in loop quantum cosmology. We find that out of the
various
proposed choices, there is only one known prescription which leads to the generic bounded behavior of these scalars. The bounds turn out
to be universal and are determined by the underlying quantum geometry. This quantization is analogous
to the so called `improved dynamics' in the isotropic loop quantum cosmology, which is also the only one to respect  the freedom of the
rescaling of
the fiducial cell at the level of effective spacetime description. Other proposed quantization prescriptions yield expansion and shear 
scalars which may not be bounded for certain 
initial conditions within the validity of effective spacetime description. These prescriptions also have a  limitation that the ``quantum
geometric effects'' can  occur at an arbitrary scale. We show that the `improved dynamics' of Kantowski-Sachs spacetime turns
out to be a unique choice in a general class of
possible quantization prescriptions, in the sense of leading to generic bounds on expansion and shear scalars and the associated physics being free
from fiducial cell
dependence. The behavior of the energy density in the `improved dynamics' reveals some interesting features. Even without considering any
details of the dynamical evolution, it is possible to rule out pancake singularities in this spacetime. The energy density is
found to be dynamically bounded. These results show that the Planck scale physics
of the loop quantized Kantowski-Sachs spacetime has key features common with the loop quantization of isotropic and Bianchi-I spacetimes.

\end{abstract}

\maketitle

\section{\bf Introduction}

Kantowski-Sachs spacetime is a homogeneous and anisotropic cosmological model which is of dual importance as it serves as both a setting to study effects of anisotropies in the  evolution of the universe and also as a description of the interior of Schwarzschild black hole in the vacuum case. This spacetime classically has a past and a future singularity, which can be an anisotropic structure  such as a barrel, cigar or a pancake, or  an isotropic point like structure depending on the initial conditions on anisotropic shear and matter\cite{collins}. At these classical singularities geodesic evolution ends, which is captured by the divergences in the expansion and shear scalars, and also of the energy density when the matter is present. The occurrence of singularities indicates that general relativity (GR) is being pushed to the  limits of its validity, and a quantum gravitational treatment of spacetime is necessary.

Though a full theory of quantum gravity is not yet available, insights on the problem of classical singularities have been gained for various spacetimes in loop quantum cosmology (LQC) in recent years \cite{as}. LQC is a quantization of  symmetry reduced spacetimes using techniques of  loop quantum gravity (LQG) which is a nonperturbative canonical quantization of gravity based on the Ashtekar variables: the SU(2) connections and the conjugate triads. The elementary variables for the quantization are the holonomies of the connection components, and the fluxes of the triads. The classical Hamiltonian constraint, the only non-trivial constraint left after symmetry reduction in the minisuperspace setting, is expressed in terms
of holonomies and fluxes and is quantized. Quantization of various isotropic models in LQC demonstrates the resolution of classical 
singularities when the spacetime curvature reaches Planck scale. The big bang and big crunch are replaced by a quantum bounce, which first 
found in the case of the spatially flat isotropic model \cite{aps1,aps2,aps3} is tied to the underlying quantum geometry and 
has been shown to be a robust phenomena through different analytical \cite{acs} and numerical investigations \cite{ps12,dgs2,squeezed}. A generalization of these results has been performed for Bianchi models \cite{chioub1,szulc_b1,awe2,b1madrid1,b1madrid2,awe3,we1,sw}, where the quantum Hamiltonian constraint also turns out to be non-singular.
An interesting feature of LQC is that for sharply peaked states which lead to a macroscopic universe at late times, it is possible to derive an effective spacetime description \cite{jw,vt,psvt}. The resulting effective dynamics has been extremely useful in not only extracting physical predictions, but also
to gain insights on the viability of various possible quantizations. In particular it has been shown that for isotropic models there is a 
unique way of quantization, the so called `improved dynamics' or the $\bar \mu$ quantization  \cite{aps3}, which results in a consistent 
ultra-violet and infra-red behavior and is free from the rescalings of the fiducial cell introduced to obtain finite integrations on the 
non-compact spatial manifold at the level of the effective spacetime description \cite{cs1,cs3}.\footnote{It should be noted that in the 
full quantum description, the independence from the fiducial cell is only approximate in the $\bar \mu$ prescription in the following 
sense. At the 
level of the physical Hilbert space,
 a mapping taking in to account fiducial rescaling mixes the superselected sectors in the quantum difference equation \cite{cs1}. 
However, for volumes much larger than the Planck volume, the rescaling invariance is recovered. This issue has been rigorously discussed in 
Refs. \cite{chiouli,cm1} where the effect of fiducial rescaling on the expectation values has been carefully studied, and rescaling 
invariance has been shown to be not satisfied at the full quantum level, but to be preserved 
 at a semi-classical level. One can also introduce an approximate rescaling invariance, such that for a suitably chosen parameters of the 
semi-classical states, the mapping does not yield a distinguishable effect \cite{cm1}. The quantization of the Kantowski-Sachs spacetime 
in the $\bar \mu$ prescription studied in this manuscript will share this caveat. However, since this issue does not arise at the level 
of the effective spacetime description which is derived using semi-classical states, the results obtained in this manuscript are 
unaffected.} Note that the fiducial cell which acts like an infra-red regulator is an arbitrary choice in the quantization procedure.  
Hence a consistent quantization prescription must yield physical predictions about observables such as expansion and shear scalars 
independent of the choice of this cell for suitable semiclassical states if the spatial topology is non-compact.

The improved dynamics quantization of the isotropic LQC results in a generic bound on the expansion scalar of the geodesics in the effective
spacetime and leads to a resolution of all possible strong singularities in the spatially flat model \cite{ps09,sv1}. These results have
also been extended to Bianchi models, where $\bar \mu$ quantization results in generic bounds on expansion and shear scalars
\cite{cs3,ps11,gs,sw}, and the resolution of strong singularities in Bianchi-I spacetime \cite{ps11}. There are other possible ways to
quantize isotropic and anisotropic models, such as the earlier quantization of isotropic models in LQC -- the $\mu_o$ quantization
\cite{abl,aps2} and the lattice refined models \cite{BCK}. In these quantization prescriptions,\footnote{Our usage of term ``quantization
prescriptions'' in loop quantization in this paper is different from an earlier work in isotropic LQC \cite{madridprescriptions}. Here
different quantum prescriptions refer to the way the area of the loops over which holonomies in
the quantum theory are constructed are constrained with respect to the minimum area gap. Whereas in Ref. \cite{madridprescriptions}, 
different quantum prescriptions were used to distinguish the quantum Hamiltonian constraints in the $\bar \mu$ quantization of isotropic 
LQC.} quantum gravitational effects can occur at arbitrarily small curvature scales and the expansion and shear scalars are not bounded in 
general \cite{cs1,cs3}. Further, the physics in these prescriptions is also not free from fiducial cell dependence at the level of efective 
dynamics. 

Loop quantization of Kantowski-Sachs spacetimes has been mostly  studied for the vacuum case \cite{ab,mod,mod2,gp1,gp2,gp3,pullin}, where the quantum Hamiltonian constraint
has been found to be non-singular. Ashtekar and Bojowald proposed a quantization of the interior of the Schwarzschild interior and concluded 
that the wavefunction of universe can be evolved across the classical central singularity pointing towards singularity resolution \cite{ab}. 
Spherically symmetric spacetimes have been  studied in the midisuperspace setting by Campiglia, Gambini, Pullin \cite{gp3,gp1,gp2}, to 
quantize Schwarzschild black hole \cite{pullin} and calculate the Hawking radiation \cite{gambini}. Though these works provide important 
insights on the quantization of black holes in LQG, it is to be noted that the  quantization prescription used in these works is analogous 
to the earlier works in isotropic LQC (the $\mu_o$ quantization) which was found to yield inconsistent physics. In particular, the loop 
quantization in these models is carried out such that the loops over which holonomies are considered have edge lengths (labelled by 
$\delta_b$ and $\delta_c$) as constant. As in the case of the $\mu_o$ quantization in LQC, the constant $\delta$
quantization of Schwarzschild interior has been shown to be dependent on the rescalings of the fiducial length $L_o$ in the $x$ direction of the $\mathbb{R} \times \mathbb{S}^2$
spatial manifold \cite{bv,dwc,chiou}. To overcome these problems, Boehmer and Vandersloot proposed a quantization prescription motivated by 
the improved dynamics in LQC \cite{bv}, which we label as $\bar \mu$ quantization in Kantowski-Sachs model. In this prescription, $\delta_b$ 
and $\delta_c$ depend on  triad components in such a way that the effective Hamiltonian constraint respects the freedom in rescaling of 
length $L_o$.\footnote{As remarked in footnote 1, strictly speaking this fiducial length independence is not present when one is 
considering volumes comparable  to the Planck volume in the full quantum description \cite{cs1,chiouli,cm1}. In the following, the fiducial 
rescaling independence will be referred to only at the effective level.} This prescription has been used to understand the phenomenology of 
the Schwarzschild interior \cite{dwc2} and has been recently used to loop quantize spherically symmetric spacetimes \cite{chiou}. It is to 
be noted that 
this prescription leads to ``quantum gravitational effects'' not only in the neighborhood of the physical singularity at 
the origin, but also at the coordinate singularity at the horizon, which points to the limitation of dealing with Schwarzschild interior 
in this setting. This problem has been noted earlier, see for eg. Ref. \cite{dwc2} where the problem with the fiducial cell at the horizon in this prescription is noted. However, note that such an issue does not arise in the presence of matter which is the focus of the present manuscript. 

In literature,  another quantization prescription inspired by the improved dynamics, which we label as the $\bar \mu'$ prescription\footnote{Our labeling of the  $\bar \mu$ and $\bar \mu'$ prescriptions in Kantowski-Sachs spacetime is opposite
to that of Ref. \cite{dwc}. This difference is important to realize to avoid any confusions about the physical implications or the limitations of these prescriptions while relating this work with Ref. \cite{dwc}.} has been proposed. In this prescription though edge lengths $\delta_b$ and $\delta_c$ are functions of the triads, problems with fiducial length rescalings persist \cite{dwc}. These prescriptions have also been analyzed for the von-Neumann stability of the quantum Hamiltonian constraints which turn out to be difference equations \cite{BCK}. It was found that $\bar \mu'$ quantization, in contrast to the $\bar \mu$ quantization, does not yield a stable evolution.

These studies indicate that if we consider fiducial length rescaling issues, $\bar \mu$ quantization in the Kantowski-Sachs spacetime is
preferred over the constant $\delta$ quantization \cite{ab} and the $\bar \mu'$ quantization prescription \cite{dwc}. However one may argue
that these issues which arise for the non-compact spatial manifold, can be avoided if the topology of the spatial manifold is  compact
($\mathbb{S}^1 \times \mathbb{S}^2$). Note that for all the models studied so far, it has been found that all three prescriptions lead to
singularity resolution. Still, little is known about the conditions under which singularity resolution occurs for the arbitrary matter.
Hence, various pertinent questions remain unanswered.  In particular, which of these quantization prescriptions promises to generically
resolve all the strong singularities\footnote{For a discussion of the strength of the singularities in LQC, see Ref. \cite{ps09}.}
within the validity of the effective spacetime description in LQC? Is it possible that in any of these quantization prescriptions,
expansion and shear scalars may not be generically bounded in effective dynamics which disfavor them
over others? Are there any other
consistent quantization prescriptions for the Kantowski-Sachs model, or is the $\bar \mu$ quantization prescription unique as in the
isotropic LQC? Finally, what is the fate of energy density if expansion and shear scalar are generically bounded? Note that in the
isotropic LQC, and the Bianchi-I model similar questions were raised in Refs. \cite{cs1,ps09,ps11}, and the answers led to $\bar \mu$
quantization as the preferred choice. It turned out to be a unique quantization prescription leading to generic bounds on expansion and shear scalars,
which were instrumental in proving the resolution of all strong singularities in the effective spacetime \cite{ps09,ps11}.

The goal of this work is to answer these questions in the effective spacetime description in LQC for Kantowski-Sachs spacetime with
minimally coupled matter. The expansion and shear scalars  are tied to the geodesic completeness of the spacetime and are independent of the fiducial
length at the classical level. We will be interested in finding the quantization prescription which promises to resolve all possible
classical singularities generically. Such a quantization prescription is expected to yield bounded behavior of these scalars. It is also
reasonable to expect, due to the underlying Planck scale quantum geometry, that in the bounce regime, depending on the approach to the
classical singularity, at least one of the scalars takes Planckian value. We find that in the effective dynamics for constant $\delta$ and
$\bar \mu'$ prescriptions, these scalars are not necessarily bounded above. In the cases where the classical singularities are resolved, it
is possible
that the expansion and shear scalars in these prescriptions can take arbitrary values in the bounce regime. In contrast,
for the $\bar \mu$ quantization prescription, we show that the expansion and shear scalars turn out to be generically bounded by universal
values in the Planck regime. It is to be noted that in the $\bar \mu$ prescription, the bounded behavior of the expansion scalar has been
mentioned earlier for the Schwarzschild interior \cite{cortez}.\footnote{We thank Alejandro Corichi for
pointing out Ref.\cite{cortez} to us.}

We find that the behavior of expansion and shear scalars in the $\bar \mu$ prescription is similar to the improved dynamics of
isotropic and Bianchi-I spacetime in LQC where the universal bounds on expansion and shear scalars were found.
Next, we address the important question of the uniqueness of the $\bar \mu$ prescription. For this we consider a general ansatz to consider
edge lengths $\delta_b$ and $\delta_c$ as functions of triads, allowing a large class of loop quantization prescriptions in the
Kantowski-Sachs spacetime. We find that demanding that the expansion and shear scalars be bounded leads to a unique choice -- the $\bar \mu$
quantization prescription. In this quantization prescription we also investigate the behavior of the energy density and find that its
potential divergence is determined only by the vanishing $g_{\Omega \Omega}$ component of the spacetime metric. This is unlike the behavior
in the classical GR, and other quantization prescriptions where divergence in energy density can occur when either of $g_{xx}$ or $g_{\Omega
\Omega}$ components vanish. An immediate consequence of this
behavior is that the pancake singularities which occur when $g_{xx}$ component of the line element approaches zero, and $g_{\Omega \Omega}$
 is finite, are forbidden. It turns out that energy density is bounded dynamically, since $g_{\Omega \Omega}$ never becomes zero and
approaches an asymptotic value. This property of $g_{\Omega\Omega}$   was first seen in the case of vacuum Kantowski-Sachs spacetime, and
turns out to be true for all perfect fluids \cite{js2}.  These results show that the $\bar \mu$
quantization in the Kantowski-Sachs spacetime is strikingly similar to the $\bar \mu$  quantization in the isotropic and Bianchi-I
spacetimes. It leads to generic bounds on the expansion and shear scalars and is independent of the rescalings of the fiducial cell at the 
effective level.

This article is organized as follows. In the next section we summarize the Kantowski-Sachs spacetime in terms of Ashtekar variables and obtain the classical equations. In Sec. III, we introduce the effective Hamiltonian constraint, and derive expressions for expansion and shear scalars for three quantization prescriptions. We discuss the boundedness of these scalars and for completeness also discuss their dependence on fiducial cell. Then, in Sec. IV, we consider a general ansatz and investigate the conditions under which a quantization prescription yields bounded behavior of expansion and shear scalars. This leads us to the  uniqueness of the $\bar \mu$ quantization prescription.  The behavior of energy density is discussed in Sec. V, which is followed by a summary of the main results.

\section{Classical Hamiltonian of Kantowski-Sachs space-time}
We consider the Kantowski-Sachs spacetime with a spatial topology of $\mathbb{R} \times \mathbb{S}^2$. Utilizing the symmetries associated with each spatial slice, the symmetry group
$\mathbb{R} \times SO(3)$, and after imposing the Gauss constraint, the Ashtekar-Barbero connection and the conjugate (densitized) triad can be expressed in the following form \cite{ab}:
\ba
A_a^i \tau_i \d x^a &=& \tilde c \tau_3 \d x + \tilde b \tau_2 \d \theta - \tilde b \tau_1 \sin\theta \d \phi + \tau_3 \cos \theta \d \phi ~,\\
\tilde E_i^a \tau_i \p_a &=& \tilde p_c \tau_3 \sin \theta \p_x + \tilde p_b \tau_2 \sin \theta \p_\theta - \tilde p_b \tau_1 \p_\phi ~,
\ea
where $\tau_i = - i \sigma_i/2$, and $\sigma_i$ are the Pauli spin matrices. The symmetry reduced triad variables are related to the metric components of the line
element,\footnote{This metric can be expressed as the one for the Schwarzschild interior by choosing $N(t)^2=\left(\frac{2m}{t}-1\right)^{-1}$ where $m$ denotes the mass of the black hole, and identifying $g_{xx}=\left(\frac{2m}{t}-1\right)$ and $g_{\Omega\Omega}=t^2$.}
\be
ds^2=-N(t)^2dt^2+g_{xx}dx^2+g_{\Omega\Omega}\left(d\theta^2+\sin^2{\theta}d\phi^2\right).
\ee
as
\ba
g_{xx}=\frac{\tilde{p_b}^2}{\tilde{p_c}}, ~~~~ \mathrm{and} ~~~~ g_{\Omega\Omega}=|\tilde{p_c}|.
\ea
The modulus sign arises because of two possible triad orientations. Without any loss of generality, we will assume the orientation to be positive throughout this analysis.
Since the spatial manifold in Kantowski-Sachs spacetime is  non-compact, we have to introduce a fiducial length along the non-compact $x$ direction. Denoting this length be $L_o$,
the symplectic structure is given by
\be
{\bf \Omega}=\frac{L_o}{2G\gamma}\left(2 \d\tilde{b}\wedge \d\tilde{p_b}+d\tilde{c}\wedge \d\tilde{p_c}\right).
\ee
Here $\gamma$ is the Barbero-Immirzi parameter whose value is fixed from the black hole entropy calculations in loop quantum gravity to be 0.2375. Since the fiducial
length can be arbitrarily rescaled, the symplectic structure depends on $L_o$. This dependence can be removed by a rescaling of the symmetry reduced triad and connection components by introducing the triads $p_b$ and $p_c$, and the connections $b$ and $c$:
\be
p_b=L_o \tilde{p_b}, \text{  }p_c=\tilde{p_c}, \text{  }b=\tilde{b}, \text{  }c=L_o \tilde{c}. \label{scaling} ~.
\ee
The non-vanishing Poisson brackets between these new variables are given by,
\ba
\left\lbrace b,p_b \right\rbrace = G \gamma, \text{            } \left\lbrace c,p_c \right\rbrace = 2G \gamma \label{poibra}.
\ea
Note that $p_b$ and $p_c$ both have dimensions of length squared, whereas $b$ and $c$ are dimensionless. 
Also note that $c$ and $p_b$ scale as $L_o$ where as other two variables are independent of the fiducial cell.

In Ashtekar variables, the Hamiltonian constraint for the Kantowski-Sachs spacetime with minimally coupled matter corresponding to an energy density $\rho_m$ can be written as
\be
\mathcal{H}_{\rm {cl}}=\frac{-N}{2G\gamma^2}\left[2bc \sqrt{p_c}+\left(b^2+\gamma^2\right)\frac{p_b}{\sqrt{p_c}}\right]\, + \, N\, 4\pi
p_b \sqrt{p_c} \rho_m, \label{ham_c}
\ee
and the physical volume of the fiducial cell is $V = 4 \pi p_b \sqrt{p_c}$. In the following, the lapse will be chosen as unity.\footnote{To make a connection with the
Schwarzschild interior, a convenient choice of lapse is $N=\frac{\gamma \sqrt{p_c}}{b}$\cite{ab}. For studies of the expansion and shear scalars and the phenomenological implications of Kantowski-Sachs spacetime with matter, the choice $N = 1$ is more useful, and is thus considered here.} Using the Hamilton's equations, for $N=1$, the dynamical equations become,
\begin{eqnarray}
\dot{p_b}&=&-G\gamma \frac{\partial \mathcal{H}_{\rm {cl}}}{\partial b}=\frac{1}{\gamma}\left(c\sqrt{p_c}+\frac{bp_b}{\sqrt{p_c}}\right)
\label{p_b}\\
\dot{p_c}&=&-2G\gamma \frac{\partial \mathcal{H}_{\rm {cl}}}{\partial c}=\frac{1}{\gamma}2b\sqrt{p_c} \label{p_c}\\
\dot{b}&=&G\gamma \frac{\partial \mathcal{H}_{\rm {cl}}}{\partial p_b}=\frac{-1}{2\gamma \sqrt{p_c}}\left( b^2+\gamma^2 \right) + 4 \pi G
\gamma \sqrt{p_c} \left( \rho_m + p_b \frac{\partial \rho_m}{\partial p_b} \right) \\
\dot{c}&=&2G\gamma \frac{\partial \mathcal{H}_{\rm {cl}}}{\partial p_c}=\frac{-1}{\gamma \sqrt{p_c}}\left(bc-\left( b^2+\gamma^2
\right)\frac{p_b}{2p_c} \right)  + 8 \pi \gamma G p_b \left( \frac{\rho_m}{2 \sqrt{p_c}} + \sqrt{p_c} \frac{\partial \rho_m}{\partial p_c}
\right) ~.
\end{eqnarray}

The vanishing of the classical Hamiltonian constraint, $\mathcal{H}_{\rm{cl}} \approx 0$, yields
\be
\f{2 b c}{\gamma^2 p_b} + \f{b^2}{\gamma^2 p_c} + \f{1}{p_c} = 8 \pi G \rho_m
\ee
which using the expressions for the directional Hubble rates $H_i=\dot{\sqrt{g_{ii}}}/\sqrt{g_{ii}}$ can be written as the Einstein's field
equation for the $0-0$ component:
\be\label{zero-zero}
2 \f{\dot{\sqrt{g_{xx}}}}{\sqrt{g_{xx}}} \f{\dot{\sqrt{g_{\Omega \Omega}}}}{\sqrt{g_{\Omega \Omega}}} + \left(\f{\dot{\sqrt{g_{\Omega
\Omega}}}}{\sqrt{g_{\Omega \Omega}}}\right)^2 + \f{1}{g_{\Omega \Omega}} = 8 \pi G \rho_m ~.
\ee
Introducing the expansion $\theta$ and the shear $\sigma^2$ of the congruence of the cosmological observers
\be
\theta = \frac{\dot{V}}{V} = \frac{\dot{p_b}}{p_b}+\frac{\dot{p_c}}{2p_c}. \label{theta}
\ee
and 
\be
\sigma^2= \frac{1}{2}\displaystyle\sum\limits_{i=1}^3 \left(H_i-\frac{1}{3}\theta\right)^2 = 
\frac{1}{3}\left(\frac{\dot{p_c}}{p_c}-\frac{\dot{p_b}}{p_b}\right)^2 \label{sigma} ~
\ee
we can rewrite eq.(\ref{zero-zero}) as
\be
\f{\theta^2}{3} - \sigma^2 + \f{1}{g_{\Omega\Omega}} = 8 \pi G \rho_m ~.
\ee

To investigate if the Kantowski-Sachs spacetime is singular, we consider the expansion and the shear scalars of the geodesics.  At a
singular region one or more of these diverge. This divergence causes the curvature invariants to blow up. To see this, we can compute the 
Ricci scalar $R$, which for the Kantowski-Sachs metric turns out to be
\be
R = 2 \f{\ddot p_b}{p_b} + \f{\ddot p_c}{p_c} + \f{2}{p_c} ~.
\ee
Using the equations for the expansion and the shear scalar, the Ricci scalar can be expressed as
\be
R = 2 \dot\theta + \f{4}{3} \theta^2 + 2 \sigma^2 + \f{2}{p_c} ~.
\ee
Thus, a divergence in $\theta$ and $\sigma^2$ signals a divergence in the Ricci scalar. For this reason, understanding the behavior of 
expansion and shear scalars is important to gain insights on not only the properties of the geodesic evolution, but it is also useful to 
understand  the behavior of curvature invariants.

The scalars,  $\theta$ and $\sigma^2$, diverge if either one or both of $\frac{\dot{p_b}}{p_b}$ and $\frac{\dot{p_c}}{p_c}$ diverge. From the Hamilton's equations of motion \eqref{p_b} and \eqref{p_c}, these ratios are,
\begin{eqnarray}
\frac{\dot{p_b}}{p_b} &=& \frac{1}{\gamma} \left( \frac{c \sqrt{p_c}}{p_b} + \frac{b}{\sqrt{p_c}} \right) \label{pbdcla} \\
\frac{\dot{p_c}}{p_c} &=& \frac{2b} {\sqrt{p_c}\gamma} ~. \label{pcdcla}
\end{eqnarray}
It is clear from  equations \eqref{pbdcla} and \eqref{pcdcla} that the expansion and shear scalars diverge as the triad components vanish, and/or the connection components diverge.
In the Kantowski-Sachs spacetime with perfect fluid as matter,  classical singularities occur at a vanishing volume. The structure of the 
singularity can be a barrel, cigar, pancake or a point \cite{collins}. For all these structures, either $p_b$ or $p_c$ vanish, causing a 
divergence in $\theta$ and $\sigma^2$.\footnote{ Note that for the vacuum Kantowski-Sachs spacetime, the expansion and shear scalars are ill 
defined 
at the horizon because of the coordinate singularity. However, $\theta^2/3 - \sigma^2$ is regular at the horizon, and can be used to 
understand the behavior of the curvature invariants. As an example, in this case, the Kretschmann scalar at the horizon can be written as 
$K_{\rm{t = 2 m}} = 12(\theta^2/3 - \sigma^2)^2$, which being finite shows that the singularity at $t=2m$ is not physical.}

At the above classical singular points, the energy density also diverges. From the vanishing of the Hamiltonian constraint $\mathcal{H}_{\mathrm{cl}}
\approx 0$, the expression for energy density becomes
\begin{equation} \label{cla_rho}
\rho_m = \frac{1}{8\pi G \gamma^2} \left[ \frac{2bc}{p_b} + \frac{b^2+\gamma^2}{p_c} \right].
\end{equation}
Thus, if either of $p_b$ or $p_c$ vanishes, $\rho_m$ grows unbounded as the physical volume approaches zero.

\section{Effective loop quantum cosmological dynamics: Comparison of different prescriptions}

Due to the underlying quantum geometry, the loop quantization of the classical Hamiltonian of the Kantowski-Sachs spacetime yields  a 
difference equation \cite{ab}. The difference equation arises due to non-local nature of the field strength of the connection in the quantum 
Hamiltonian constraint which is expressed in terms of holonomies of connection components over closed loops. The action of the holonomy 
operators on the triad states is discrete, leading to a discrete quantum Hamiltonian constraint which is non-singular.\footnote{In 
principle, there can also be inverse triad modifications in the quantum Hamiltonian constraint. However, such modifications can not be 
consistently defined for spatially non-compact manifolds not only at the full quantum level but even in the effective Hamiltonian 
description, due to the dependence on the fiducial length. As discussed earlier in footnotes 1 and 2, note that under the rescaling of the 
fiducial cell, the 
invariance of the quantum theory does not hold when volume is comparable to the Planck volume. However, at larger volumes in the 
quantum theory, and also in the approximation of the validity of the effective dynamics $\bar \mu$ prescription preserves rescaling under 
fiducial cell. In this analysis, we do not consider inverse triad modifications. However, these can be consistenly included if 
 the spatial topology is compact, and conclusions reached in this manuscript remain unaffected in this case. Further, it is also possible 
to get rid of terms depending on inverse triad using a suitable choice of lapse.}
The resulting quantum dynamics can be captured using an effective Hamiltonian constraint derived using the geometrical formulation of
quantum mechanics \cite{aa_ts}. Here one treats the Hilbert space as a quantum phase space and seeks an embedding of the finite
dimensional classical phase space into it. For the isotropic and homogeneous models in LQC, such a suitable embedding has been found using
sharply peaked states which probe volumes larger than the Planck volume \cite{vt,psvt}.  For these models, the dynamics from the quantum
difference equation and the effective Hamiltonian turn out to be in an excellent agreement for states which correspond to a classical
macroscopic universe at late times.
Recent numerical investigations show that the departures between the effective spacetime description and the quantum dynamics are negligible unless one consider states which
correspond to highly quantum spacetimes, such as states which are widely spread or are highly squeezed and non-Gaussian, or those which do 
not lead to a classical universe at late times \cite{dgs2,squeezed}. 
Though the effective Hamiltonian constraint has not been derived for the anisotropic spacetimes in LQC using the above embedding approach,
an expression for it has been obtained by replacing $b$ with $\frac{\sin{b \delta_b}}{\delta_b}$ and $c$ with $\frac{\sin{c\delta_c}}{\delta_c}$ in \eqref{ham_c}, where
$\delta_b$ and $\delta_c$ are the edge lengths of the holonomies \cite{dwc,bv}. Following this procedure for the case of the
loop quantization of the vacuum Bianchi-I spacetime, the resulting effective Hamiltonian dynamics turns out to be in excellent agreement with the underlying quantum
evolution \cite{madrid_b1}. In the following we will assume that the effective Hamiltonian constraint for the Kantowski-Sachs
spacetime as obtained from the above polymerization of the connection components, and assume it to be valid for all values of triads. For a
general choice of $\delta_b$ and $\delta_c$, the effective Hamiltonian constraint for the Kantowski-Sachs model with matter is given as
\cite{dwc,bv}:
\be
\mathcal{H}=\frac{-N}{2G\gamma^2}\left[2\frac{\sin{(b\delta_b)}}{\delta_b}\frac{\sin{(c\delta_c)}}{\delta_c}\sqrt{p_c}+\left(\frac{\sin^2{(b\delta_b)}}{\delta_b^2}+\gamma^2\right)\frac{p_b}{\sqrt{p_c}}\right]+N4\pi p_b \sqrt{p_c} \rho_m. \label{ham_gen}
\ee
Note that \eqref{ham_gen} goes to the classical Hamiltonian \eqref{ham_c} in the limit $\delta_b \rightarrow 0$ and $\delta_c \rightarrow 0$. However, due to the existence of minimum area gap in LQG, in the quantum theory, one shrinks the loops to the minimum finite area. Different choices of the way holonomy loops are constructed and shrunk lead to different $\delta_b$ and $\delta_c$, and different properties of the quantum Hamiltonian constraint. We will identify these choices as different prescriptions to quantize the theory, which lead to different functional forms of $\delta_b$ and $\delta_c$ in the polymerization of the
connection, and hence result in different effective Hamiltonian constraints. This is analogous to the situation in the quantization of
isotropic spacetimes in LQC, where the older quantization was based on constant $\delta$ (the so called $\mu_o$ quantization
\cite{abl,aps2}), and improved quantization is based on a $\delta$ which is function of isotropic triad $\delta \propto 1/\sqrt{p}$ (the so
called $\bar \mu$ quantization \cite{aps3}).  As in the isotropic case, the physics obtained from the theory is dependent on these holonomy edge lengths and hence they have to be chosen carefully. This can be further seen by noting that  $\sin{(b\delta_b)}$ and $\sin(c\delta_c)$ in (\ref{ham_gen}) can be expanded in infinite series as $b\delta_b-\frac{b^3\delta_b^3}{3!} + ...$ and  $c\delta_c-\frac{c^3\delta_c^3}{3!} + ...$ \, . Hence it is required that $b\delta_b$ and $c\delta_c$ should be independent of fiducial length. Else different terms of the expansion will have different powers of $L_o$ and any calculation based on this Hamiltonian will yield results which are sensitive to the choice of $L_o$. Of the possible choices of holonomy edge lengths that can be motivated, we have to choose the one that gives a mathematically consistent theory which renders the physical scalars such as expansion and shear scalars independent of the choice of fiducial length, as in classical GR.  There are three proposed
prescriptions in LQC literature for the choice of holonomy edge-lengths in the Kantowski-Sachs model: the
constant $\delta$ \cite{ab}, the $\bar \mu$ (or the `improved dynamics') prescription \cite{bv}, and the $\bar \mu'$ (inspired from the improved dynamics) quantization prescriptions \cite{dwc}. Due to their similarities with the notation of the isotropic model, we will label the effective Hamiltonian constraint for
constant $\delta$ with $\mu_o$. The effective Hamiltonians for `improved dynamics' inspired prescription will be labelled by $\bar \mu'$, and that of  `improved dynamics' prescription with $\bar \mu$.

\subsection{Constant $\delta$ prescription}

The simplest choice of $\delta's$ is to choose them as constant. The resulting effective Hamiltonian constraint then corresponds
to the loop quantization
of Kantowski-Sachs spacetime where the holonomy considered over the loop in $x-\theta$ plane, and the loop in the $\theta-\phi$ plane  has
minimum area with respect to the fiducial metric fixed by the minimum area eigenvalue $\Delta$ in LQG: $\Delta = 4 \sqrt{3} \pi \gamma
\lp^2$. In the quantization of the Schwarzschild interior proposed in Ref. \cite{ab}, the $\delta's$ were chosen equal\footnote{Since \cite{ab} was using an area gap of $\Delta = 2 \sqrt{3} \pi \gamma
\lp^2$, the corresponding holonomy edge lengths were $2\sqrt{3}$. For $\Delta = 4 \sqrt{3} \pi \gamma$, edge lengths should be $4
\sqrt{3}$.} $\delta_b =
\delta_c = 4 \sqrt{3}$. Loop quantization with constant $\delta_b$ and $\delta_c$ is also considered in various other works on the
loop quantization of black hole spacetimes \cite{gambini,gp1,gp2}, and is analogous to the $\mu_o$ quantization in the isotropic LQC
\cite{abl,aps2}. Here we will assume the same prescription in the presence of matter.

The resulting effective Hamiltonian constraint for $N=1$ with minimally coupled matter is:
\be
\mathcal{H}_{\mu_0}=\frac{-1}{2G\gamma^2}\left[2\frac{\sin{(b\delta_b)}}{\delta_b}\frac{\sin{(c\delta_c)}}{\delta_c}\sqrt{p_c}+\left(\frac{\sin^2{(b\delta_b)}}{\delta_b^2}+\gamma^2\right)\frac{p_b}{\sqrt{p_c}}\right]+4\pi p_b \sqrt{p_c} \rho_m. \label{ham_cd}
\ee
Using the Hamilton's equations, the equations of motion for the triads are
\ba
\dot{p_b} &=&-G\gamma \frac{\partial \mathcal{H}_{\mu_0}}{\partial b}= \frac{1}{\gamma} \left(\cos{(b \delta_b) \frac{\sin{(c \delta_c)}}{\delta_c}} \sqrt{p_c} + \frac{\sin{(b \delta_b)}\cos{(b \delta_b)}}{\delta_b}\frac{p_b}{\sqrt{p_c}}\right), \\
\dot{p_c} &=&-2G\gamma \frac{\partial \mathcal{H}_{\mu_0}}{\partial c}= \frac{2}{\gamma}\cos{(c \delta_c) \frac{\sin{(b \delta_b)}}{\delta_b}} \sqrt{p_c}.
\ea
From these one can find the expressions for expansion\footnote{The expressions for $\theta$ in three prescriptions studied in this section
were also obtained for the Schwarzschild interior  in
Ref.\cite{cortez},  however no physical implications were studied except for noticing the bounded behavior in the case of $\bar \mu$
prescription.}   and shear
scalars for Kantowski-Sachs spacetime with matter as follows,
\ba
\theta &=& \frac{1}{\gamma} \left( \frac{\sqrt{p_c}\cos{(b\delta_b)\sin{(c \delta_c)}}}{p_b\delta_c}+ \frac{\sin{(b\delta_b)}}{\sqrt{p_c}\delta_b}\left(\cos{(b\delta_b)}+\cos{(c\delta_c)}\right) \right) \label{thetamuo} \\
\sigma^2 &=& \frac{1}{3 \gamma^2} \left( \left(2 \cos{(c \delta_c)}-\cos{(b \delta_b)}\right)\frac{\sin{(b \delta_b)}}{\delta_b \sqrt{p_c}} - \frac{\cos{(b\delta_b)}\sin(c\delta_c)}{\delta_c}\frac{\sqrt{p_c}}{p_b} \right)^2 \label{shearmuo} ~.
\ea
It is clear from the above expressions that the expansion and shear scalars are unbounded and blow up as $p_b$ or $p_c$ approach zero, precisely as in
the classical Kantowski-Sachs spacetime if the effective spacetime description is assumed to be valid for all values of triads.
Note that the effective spacetime description is expected to breakdown in the regime when the volume of the spacetime is less than Planck
volume \cite{dgs2}. Hence, in this quantization prescription there are no generic bounds on the expansion and shear scalars within the
expected validity of effective dynamics. Even if
one considers a specific matter model which results in a singularity resolution and a bounce of the mean volume, 
the dependence of $\theta$ and $\sigma^2$ on the triads shows that these scalars may not necessarily take
Planckian values in the bounce regime. The spacetime curvature in the bounce regime can in principle be extremely small in this effective
dynamics. Note that the maximum value of expansion (\ref{thetamuo}) and shear scalars (\ref{shearmuo}) depends on the values of $p_b$ and
$p_c$. Since the values of triads at the bounce can be made arbitrarily large or small by the choice of initial conditions and the matter
content, the maximum values of expansion and shear scalars,  reached near the bounce, can hence take arbitrary values.
This problem is analogous to the dependence of energy density at the bounce on the momentum of the scalar field or the triad in the $\mu_o$ quantization of isotropic LQC. There too by choosing different initial conditions it is possible to obtain ``quantum bounce'' at arbitrarily small spacetime curvature.

Let us now consider the issue of fiducial cell dependence for this prescription. Since $\delta_b = \delta_c = 4 \sqrt{3}$,  they are independent of the rescaling under the fiducial length $L_o$. However, since $c$ is
proportional to $L_o$, therefore $c \delta_c$ depends on the fiducial length $L_o$. Due to this reason, the resulting physics from the
effective
Hamiltonian constraint (\ref{ham_cd}), in particular the expressions for expansion and shear scalars, unlike in the classical theory, are not independent of the fiducial length rescaling.  Again this problem of constant $\delta$ prescription in the Kantowski-Sachs spacetime is analogous to the one for the $\mu_o$
quantization of the isotropic LQC, where the resulting physical predictions such as the scale at which the quantum bounce occurs and the
infra-red behavior depend on the fiducial volume of the fiducial cell \cite{aps2,cs1}. This problem is tied to the dependence of the expansion and triad scalars in this quantization
prescription on triads as discussed above. Since $p_b$ can be rescaled arbitrarily by rescaling $L_o$,  the curvature scale in the bounce regime inevitably depends on the fiducial length $L_o$ and hence can take arbitrary values.

In conclusion, we find that constant $\delta$ quantization prescription does not  provide a generic bounded behavior of expansion and shear scalars. Further,  it is possible to obtain ``quantum gravitational effects,'' originating from the trigonometric functions in eq.(\ref{ham_cd}), at any arbitrary scale.

\subsection{An `improved dynamics inspired' prescription}
For the isotropic models in LQC, the problems with constant $\delta$ (i.e. $\mu_o$) quantization were overcome in the improved dynamics (the $\bar \mu$ quantization) \cite{aps3},
where
$\bar \mu$ is related to the
isotropic triad as $\bar \mu = \Delta/\sqrt{p}$ \cite{aps3}. This quantization turns out to be independent of the various problems of the
$\mu_o$ quantization, and is also the unique prescription for the quantization of isotropic models in which physical predictions
are free of the dependence on the fiducial cell in the effective spacetime description \cite{cs1}. Motivated by the success of $\bar \mu$
quantization, a different prescription for the choice of $\delta_b$ and $\delta_c$ for Kantowski-Sachs model has been considered \cite{dwc}, where 
\be
\delta_b=\sqrt{\frac{\Delta}{p_b}} , ~~ {\rm{and}} ~~ \delta_c=\sqrt{\frac{\Delta}{p_c}} ~.
\ee
We note that this choice for $\delta's$ is also motivated from the lattice refinement scheme \cite{BCK}. 

The effective Hamiltonian constraint for this quantization becomes: 
\be
\mathcal{H}_{\bar\mu'}=\frac{-1}{2G\gamma^2\Delta}\left[2\sin(b\delta_b)\sin(c\delta_c)p_c\sqrt{p_b}+\left(\sin^2(b\delta_b)p_b+\gamma^2\Delta\right)\frac{p_b}{\sqrt{p_c}}\right]+4\pi p_b\sqrt{p_c}\rho_m ~.
\ee
As we noted above, for the effective Hamiltonian constraint to yield a consistent physics, the argument of trigonometric functions should be independent of the fiducial length.
However since $b$ is independent of $L_o$ and $p_b$ is proportional to $L_o$,   $b \delta_b = b \sqrt{\frac{\Delta}{p_b}}$ depends on
fiducial length. Similarly $c \delta_c$ also depends on the fiducial length. This clearly shows that this quantization is unsuitable for
Kantowski-Sachs spacetime because the resulting physical implications will be sensitive to the fiducial length $L_o$.   

The equations of motion for the triads in this quantization are
\ba
\dot{p_b}&=&-G\gamma \frac{\partial \mathcal{H}_{\bar \mu'}}{\partial b}=\frac{\cos(b \delta_b)}{\gamma \sqrt{\Delta}}\left(p_c \sin(c \delta_c)+p_b\sqrt{\frac{p_b}{p_c}}\sin(b\delta_b)\right) \\
\dot{p_c}&=&-2G\gamma \frac{\partial \mathcal{H}_{\bar\mu'}}{\partial c}=\frac{2}{\gamma \sqrt{\Delta}}\sqrt{p_bp_c}\sin(b \delta_b)\cos(c \delta_c),
\ea
using which the expansion and shear scalars turn out to be as follows:
\ba
\theta &=& \frac{1}{\gamma \sqrt{\Delta}} \left[\frac{p_c}{p_b} \cos(b \delta_b) \sin(c \delta_c) +
\sqrt{\frac{p_b}{p_c}}\sin(b\delta_b)\left(\cos(b\delta_b)+\cos(c\delta_c)\right) \right], \label{thetaimpinspired} \\
\sigma^2 &=& \frac{1}{3\gamma^2 \Delta} \left[\frac{p_c}{p_b} \cos(b \delta_b) \sin(c \delta_c) + \sqrt{\frac{p_b}{p_c}}\sin(b\delta_b)\left(\cos(b\delta_b)-2\cos(c\delta_c)\right) \right]^2 ~.
\ea
We see that the $\bar\mu'$ quantization has the same problem as the constant $\delta$ quantization as far as the divergence of $\theta$ and $\sigma^2$ is concerned. These scalars can potentially diverge for $p_b \rightarrow 0$, $p_b\rightarrow \infty$, $p_c \rightarrow 0$ or $p_c
\rightarrow \infty$.

As in the constant $\delta$ quantization prescription, even if the singularities are resolved,
the curvature scale associated with singularity resolution can be arbitrarily small and depends on the initial conditions.
Also remembering that it has spurious dependency on the fiducial length we are led
to the conclusion that $\bar\mu'$ quantization is not apt for Kantowski-Sachs spacetime. The results that constant $\delta$ and $\bar\mu'$
quantizations do not yield necessarily consistent physics is in accordance with a similar study in FRW model in LQC \cite{cs1}. As remarked
earlier, problems of this prescription have also been noted in the context of the von-Neumann stability analysis of the resulting quantum
Hamiltonian constraint \cite{BCK}.\footnote{For different prescriptions, the problems in the effective dynamics and the numerical
instability of the quantum difference equation in the corresponding quantization run in parallel. See Ref. \cite{ps12} for a discussion of
these issues in different quantizations in LQC.}

\subsection{`Improved Dynamics' prescription}

The improved dynamics prescription is based on noting that the field strength of the  Ashtekar-Barbero connection should be computed by 
considering holonomies around the loop whose minimum  area with respect to the physical metric is fixed by the minimum area eigenvalue 
($\Delta$) in LQG. This is in contrast to the constant $\delta$ prescription where the minimum area with respect to the fiducial metric was 
fixed with respect to the underlying quantum geometry. 
In this scheme we obtain the holonomy edge lengths as \cite{bv}:
\be
\delta_b=\sqrt{\frac{\Delta}{p_c}}, \text{     } \delta_c=\frac{\sqrt{\Delta p_c}}{p_b} \label{imdy}.
\ee
Now the effective Hamiltonian \eqref{ham_gen} becomes,
\be
\mathcal{H}_{\bar{\mu}}=\frac{-p_b\sqrt{p_c}}{2G \gamma^2 \Delta} \left[2\sin{(b\delta_b)}\sin{(c\delta_c)}+\sin^2{(b\delta_b)}+\frac{\gamma^2 \Delta}{p_c}\right]+4\pi p_b \sqrt{p_c} \rho_m \label{ham}.
\ee
Before we proceed further, we note an important property of this effective Hamiltonian not shared by $\mathcal{H}_{\mu_o}$ and $\mathcal{H}_{\bar \mu'}$. Due to the scaling properties of $b,c,p_b$ and $p_c$, $b \delta_b$ and $c \delta_c$ are invariant under the change of the fiducial length $L_o$. Thus
$\sin{(b\delta_b)}$ and $\sin(c\delta_c)$ are independent of fiducial length. Due to this reason, we expect that the physical predictions concerning scalars such as expansion and shear scalars will be independent of $L_o$ in this prescription, as in the classical theory.

The evolution equations for triads and cotriads turn out to be as follows:
\ba
\dot{p_b}&=&-G\gamma \frac{\partial \mathcal{H}_{\bar{\mu}}}{\partial b}=\frac{p_b \cos{(b\delta_b)}}{\gamma \sqrt{\Delta}} \left( \sin{(c\delta_c)} + \sin{(b\delta_b)} \right), \label{pb-eff}\\
\dot{p_c}&=&-2G\gamma \frac{\partial \mathcal{H}_{\bar{\mu}}}{\partial c}=\frac{2p_c}{\gamma \sqrt{\Delta}} \sin{(b\delta_b)} \cos{(c\delta_c)} \label{pc-eff} \\
\ea
Using \eqref{theta}, \eqref{pb-eff} and \eqref{pc-eff}, we obtain the following expression for the expansion scalar,
\ba
\theta &=& \frac{1}{\gamma \sqrt{\Delta}} \left(  \sin{(b\delta_b)} \cos{(c\delta_c)}+\cos{(b\delta_b)} \sin{(c\delta_c)} + \sin{(b\delta_b)} \cos{(b \delta_b)} \right).
\ea
Unlike the case of ${\mathcal H}_{\mu_o}$ and $\mathcal{H}_{\bar \mu'}$, the expansion scalar turns out to be independent of the fiducial length $L_o$, and is generically bounded above by a universal value:
\be
|\theta| \leq \frac{3}{2 \gamma \sqrt{\Delta}} \approx \frac{2.78}{\lp}.
\ee
Similarly for the shear scalar, using \eqref{sigma}, \eqref{pb-eff} and \eqref{pc-eff}, we get
\ba
\sigma^2 = \frac{1}{3\gamma^2 \Delta} \left(2\sin{(b\delta_b)}\cos{(c\delta_c)}-\cos{(b\delta_b)}\left(\sin{(c\delta_c)}+\sin{(b\delta_b)} \right) \right)^2 ~.
\ea
As for the expansion scalar, $\sigma^2$ turns out to be independent of $L_o$ and has a universal maximum:
\be
\sigma^2 \leq \frac{5.76}{\lp^2}.
\ee
Hence both shear and expansion scalars are bounded above in this quantization prescription of the Kantowski-Sachs spacetime. Unlike constant $\delta$ and $\bar \mu'$ quantization prescriptions, the expansion and shear scalars take Planckian values in the bounce regime and curvature scale associated with singularity resolution does not depend on the initial conditions. Note that for the improved dynamics prescription, similar properties of expansion and shear scalar were earlier found for the isotropic model \cite{ps09} and the Bianchi models \cite{cs3,ps11,gs,sw}. In the isotropic and Bianchi-I model, using the boundedness properties of expansion and shear scalars it was found that strong singularities are generically resolved in the effective spacetime description \cite{ps09,ps11}.\footnote{These results have also been extended to the effective description of the hybrid quantization of Gowdy models \cite{gowdy}.} Above results provide a strong evidence that strong singularities may be generically absent in
this quantization of Kantowski-Sachs spacetime.

\section{Uniqueness of $\bar \mu$ prescription}
In the previous section, we found that of the three proposed quantization prescriptions for the Kantowski-Sachs spacetime in LQC, only the
the $\bar \mu$ effective Hamiltonian leads to consistent physics and results in generic bounds on expansion and shear scalars.
In this section we pose the question whether $\bar \mu$ quantization is the only possible choice for which the expansion and shear
scalars are generically bounded singularity resolution in the Kantowski-Sachs spacetime? A similar question was posed in the isotropic
models in LQC, where the answer turned out to be positive \cite{cs1,cs3}. We will see that in the Kantowski-Sachs spacetime, under the
assumption that $\delta_b$ and $\delta_c$ have a general form given in eq.(\ref{deltainv}), the answer also turns to be  in an
affirmative in the effective spacetime description.

We start with the effective LQC Hamiltonian \eqref{ham_gen}, where the holonomy edge lengths
$\delta_b$ and $\delta_c$ are any general functions of the
triads. Then the Hamilton's equations lead to the following expressions for shear and expansion scalars.
\be \label{theta_gen}
\theta = \frac{1}{\gamma} \left( \frac{\sqrt{p_c}\cos{(b\delta_b(p_b,p_c))\sin{(c \delta_c(p_b,p_c))}}}{p_b\delta_c(p_b,p_c)}+
\frac{\sin{(b\delta_b(p_b,p_c))}}{\sqrt{p_c}\delta_b(p_b,p_c)}\left(\cos{(b\delta_b(p_b,p_c))}+\cos{(c\delta_c(p_b,p_c))}\right) \right)
\ee
\ba \label{sigma_gen}
\sigma^2 &=& \frac{1}{3 \gamma^2} \bigg[ \left(2 \cos{(c \delta_c(p_b,p_c)}-\cos{(b \delta_b(p_b,p_c))}\right)\frac{\sin{(b
\delta_b(p_b,p_c))}}{\delta_b(p_b,p_c) \sqrt{p_c}} \nonumber \\ &-&
\frac{\cos{(b\delta_b(p_b,p_c))}\sin(c\delta_c(p_b,p_c))}{\delta_c(p_b,p_c)}\frac{\sqrt{p_c}}{p_b} \bigg]^2 ~.
\ea
We now find what general choices of $\delta_b(p_b,p_c)$,$\delta_c(p_b,p_c)$ yield a bound on expansion and shear scalars.
These scalars become unbounded when either an inverse power of a triad blows up as that triad tends to zero or when a positive
power of triad blows up as that triad tend to infinity. In eqs. \eqref{theta_gen} and \eqref{sigma_gen}, the trigonometric factors are
always bounded and hence the terms that will decide the boundedness of the expansion and shear scalars are
\be
T_b=\frac{1}{\sqrt{p_c}\delta_b(p_b,p_c)} ~~~ {\rm{and}} ~~~
T_c=\frac{\sqrt{p_c}}{p_b\delta_c(p_b,p_c)} . \label{T1T2}
\ee
Then the task at hand reduces to finding general functions of triads which when chosen as the holonomy edge lengths, give an upper bound on
$T_c$ and $T_b$. To this end we make an assumption that $\delta_b$ and $\delta_c$ are functions of $p_b$ and $p_c$ such that one can express
their inverses as
\be
\delta_b^{-1}=\sum B_{ij}p_b^{m_i}p_c^{n_j}, \quad \delta_c^{-1}=\sum C_{ij}p_b^{m_i}p_c^{n_j} \label{deltainv},
\ee
where $m_i,n_j \in \mathbb{R}$. This ansatz includes all the three choices of $\delta_b$ and $\delta_c$ discussed in Sec. III, but is more
general. Using (\ref{deltainv}), one can write \eqref{T1T2} as
\ba
T_c=\sum C_{ij} p_b^{m_i-1}p_c^{n_j+1/2} \label{t1} ~, \\
T_b=\sum B_{ij} p_b^{m_i} p_c^{n_j-1/2} \label{t2} ~.
\ea
We now require that if $\theta$ and $\sigma^2$ have to be bounded then $T_c$ and $T_b$ should not diverge as triads tend to zero or
infinity. This is possible only if $m_i$ and $n_j$ in \eqref{t1} and \eqref{t2} satisfy certain constraints. We find that these constraints
only allow $\delta_b \propto (p_c)^{-1/2}$ and $\delta_c \propto p_c^{1/2}/p_b$, the same as in the $\bar \mu$ quantization (\ref{imdy}).

First let us take a closer look at \eqref{t1} from which we wish to obtain constraints on $\delta_c$. Keeping $p_c$ as nondiverging and
nonvanishing, one can obtain bounds on values of $m_i$, the powers of $p_b$ with nonzero coefficients. As $p_b \rightarrow 0$, for each term
in $T_c$ to be nondiverging, they should all have a non-negative power of $p_b$. Thus, for any nonzero $C_{ij}$, $m_i \geq 1$.
Also, as $p_b \rightarrow \infty$, any positive power of $p_b$ diverges. Hence for $T_c$ to be bounded, for any nonzero $C_{ij}$, $m_{i}
\leq
1$. Therefore, the only possible value for $m_i$ that leaves $T_c$ bounded for $p_b \rightarrow 0$ and $p_b \rightarrow \infty$ is $m_i=1$.
Similarly, to find the allowed values for $n_j$, we study
the behavior of $T_c$ as $p_c$ goes to zero and infinity for a finite nonzero value of $p_b$. It is clear that positive powers of $p_c$
will result in a divergence of $T_c$ as $p_c \rightarrow \infty$ where as negative powers will result in a divergence when $p_c \rightarrow
0$. This implies that the only choice of $n_j$ that leaves $T_c$ bounded for the whole range of $p_c$ is $n_j=-1/2$. Finally, we consider
the case of both the triads simultaneously approaching one of the extreme values - zero or infinity. For $m_i
=1$ and $n_j=-1/2$, from  \eqref{t1} it can be seen that $T_c$ is independent of triads i.e, it is just a constant. Hence for both the triads
simultaneously approaching an extreme value, $T_c$ remains bounded. For any other choice of $m_i$ or $n_j$, $T_c$ can diverge, causing a
divergence in the expansion and shear scalars.

Repeating the same analysis, for $T_b$ in \eqref{t2}, it can be seen
that the only values  of $m_i$ and $n_j$ which keep $T_b$ bounded for the whole domain of $p_b$ and $p_c$ are $m_i=0$ and
$p_c=1/2$. Thus from \eqref{deltainv} it can be seen that the only choice of $\delta's$  which keeps $\theta$ and $\sigma^2$ 
bounded throughout the entire domain of triads correspond to
\ba
\delta_c \propto \frac{\sqrt{p_c}}{p_b}, ~~~
\delta_b \propto \frac{1}{\sqrt{p_c}} ~.
\ea
These are precisely the functional dependencies of the holonomy edge lengths on these triads in the `improved dynamics' prescription. (\ref{ham}). Thus, for the
general ansatz (\ref{deltainv}) we find that the only possible choices of $\delta_b$ and $\delta_c$ which result in bounded expansion and
shear scalars for the geodesics in the effective dynamics correspond to $\bar \mu$ prescription. It is important to stress that we found the uniqueness of $\bar \mu$ quantization prescription by only demanding that the expansion and shear scalars be bounded, and our argument is not tied to requirements based on fiducial cell rescaling freedom or to the topology of the spatial manifold. But, it is rather interesting that the  prescription which results in generic bounds on scalars is the one which is also free from the freedom under rescalings of the fiducial cell. It is staightforward to see that requiring $b \delta_b$ and $c \delta_c$ to be independent of fiducial length $L_o$, and assuming that $\delta_b$ and $\delta_c$  are constructed from the triads $p_b$ and $p_c$, one is led to the $\bar \mu$ prescription.

In the above analysis we have seen that by requiring that the expansion and shear scalars be always bounded, one can find the exact
dependence of
$\delta_b$ and $\delta_c$ on the triads. The same functional forms of  $\delta_b$ and $\delta_c$ can be obtained from
an independent physical
motivation. Note that holonomy corrections in the effective Hamiltonian arise from the field strength of the connection components $b$
and $c$, where one has to take the holonomies around
closed loops with edge lengths determined by $\delta_b$ and $\delta_c$. To compute the field strength, the loops over which the holonomies
are considered are shrunk to the minimum area eigenvalue in LQG. 
One could in principle form loops from holonomies with constant edgelengths
$\delta_b$, $\delta_c$ or as in the $\bar{\mu}'$ scheme, where $\delta_b=\sqrt{\frac{\Delta}{p_b}}$ and
$\delta_c=\sqrt{\frac{\Delta}{p_c}}$. But loops with such edge lengths do not have physical area matching the minimum area gap from LQG. The
constant $\delta$ quantization takes the holonomy loops
to have constant fiducial area, but not the physical area. However, fiducial area is not independent of rescaling of fiducial length and
thus is not a physical quantity. In this quantization, a loop with edges of length $\delta_b$ along $\theta$ and $\phi$ directions will
have a physical area $\delta_b^2p_c$.\footnote{It is straightforward to see that the same conclusion is reached or the loop in
$x-\theta$ plane.}
This area is clearly dependent of the triad and can even vanish as $p_c \rightarrow 0$, thus becoming smaller than the
minimum area eigenvalue of LQG. Similarly,
in $\bar{\mu}'$ quantization, the area of a loop with edge $\delta_b$ each along $\theta$ and $\phi$ directions will be $\frac{\Delta
p_c}{p_b}$. Once again this area is not constant and can go below the minimum area gap of LQG if $p_c/p_b$ becomes less than unity. In
contrast the
loops constructed in the improved dynamics with $\delta_b = \sqrt{\Delta/p_c}$ and $\delta_c = \sqrt{\Delta p_c}/p_b$ in $x-\theta$ and
$\theta-\phi$ planes have a physical area $\Delta$, which is same as the minimum area gap. Thus, this argument further supports the
improved dynamics or the $\bar \mu$ prescription for the Kantowski-Sachs spacetime.

\section{Energy density in the `improved dynamics'}
We have so far seen that out of various possible quantization prescriptions, the $\bar \mu$ prescription for the Kantowski-Sachs spacetime 
is the only one which results in bounded expansion and shear scalars for all the values of triads. Also, the resulting physics turns out to 
be independent of the rescalings under fiducial length at the level of effective spacetime description. In this sense, this is the 
preferred choice for the loop quantization in the Kantowski-Sachs model. We now investigate the issue of the boundedness of the energy 
density in this prescription. It will be useful to recall some features of classical singularities in this context. In classical GR, 
approach to singularities in the Kantowski-Sachs spacetime is accompanied by a divergence in the energy density for perfect fluids when the 
volume vanishes \cite{collins}. The nature of the singularity -- whether it is isotropic or anisotropic depends on the equation of state of 
matter. Apart from the  isotropic or the point like singularity, cigar, pancake and barrel singularities can also
form in the
classical
Kantowski-Sachs spacetime. For the point singularity both  $g_{xx}$ and $g_{\Omega \Omega}$ vanish, for the cigar singularity  $g_{xx} \rightarrow \infty$ and $g_{\Omega \Omega} \rightarrow 0$, for the barrel singularity $g_{xx}$ approaches a finite value and $g_{\Omega \Omega} \rightarrow 0$, and for the pancake singularity $g_{xx}$ vanishes and $g_{\Omega \Omega}$ approaches a finite value. In terms of the triad components, for point, cigar and barrel singularities both $p_b$ and $p_c$ vanish. However, the pancake singularity occurs at a finite value of $p_c$, with $p_b$ vanishing.

\begin{figure}[]
\includegraphics[scale=.5]{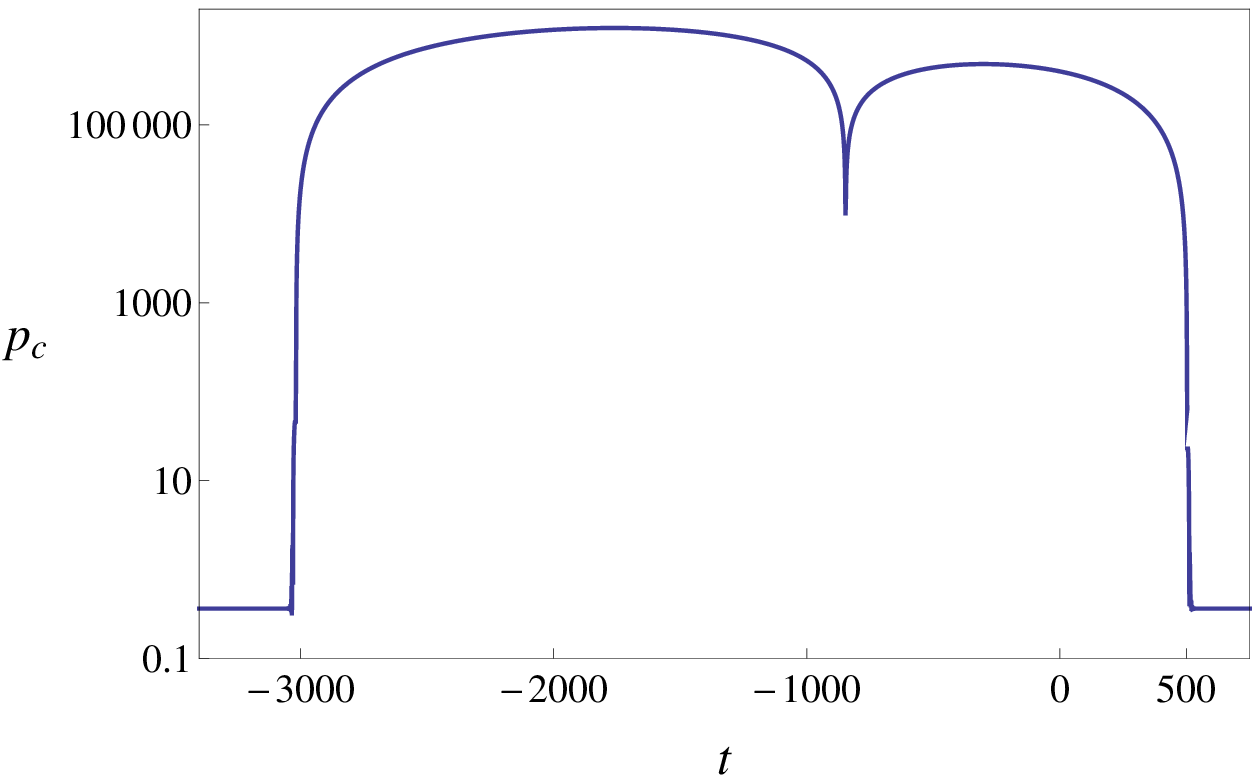}
\hskip1.5cm
\includegraphics[scale=.5]{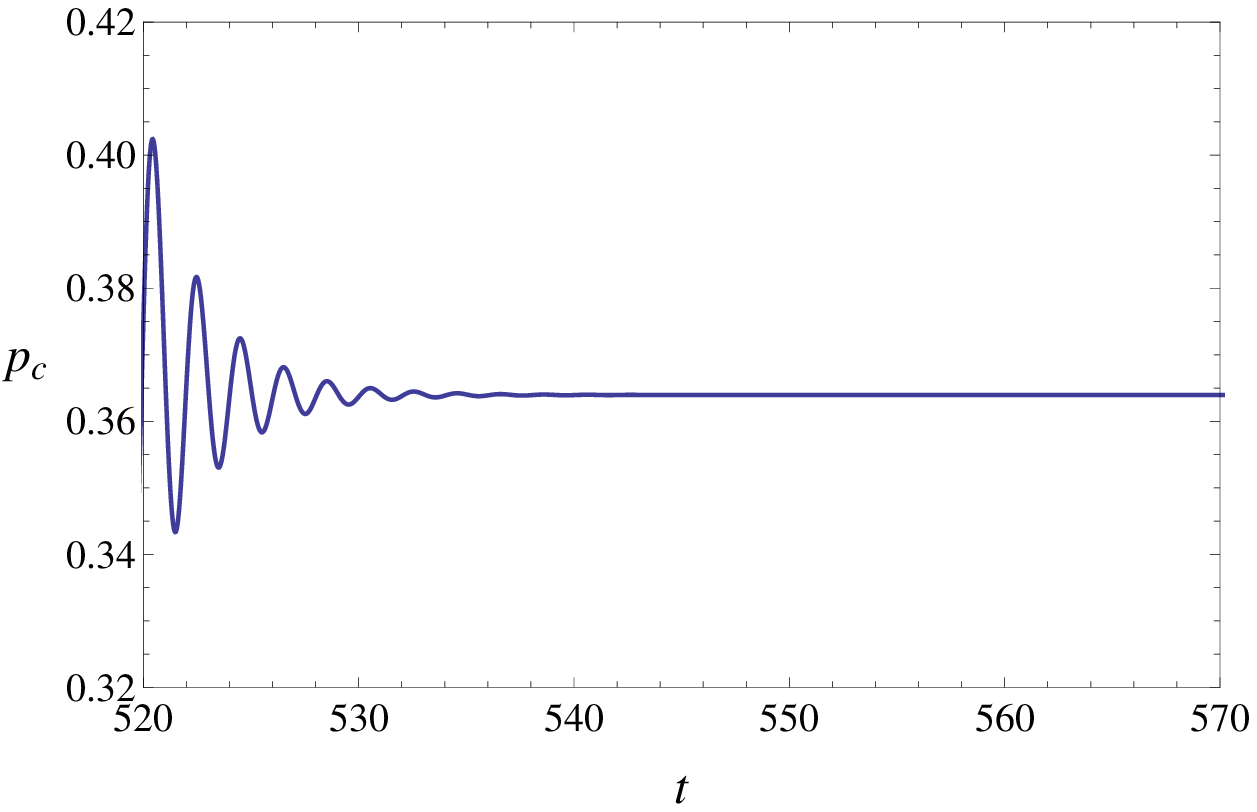}
\caption{Evolution of the triad component $p_c$ is shown as  a function of proper time for the massless scalar field evolution in the $\bar
\mu$ effective dynamics. The initial conditions are $p_b(0)=5\times 10^5$, $b(0)=-0.1$, $p_c(0)=4\times 10^5$, $c(0)=0.16$ (all in Planck
units). Initial value of energy density is obtained by solving the Hamiltonian constraint. We see that the classical singularity is avoided,
and  $p_c$ is non-zero in the entire evolution. Asymptotic approach of $p_c$ to a finite value is also shown. A similar plot is obtained 
for the vacuum case, where it was shown that some cycles of classical phases appear before $p_c$ reaches Planck regime \cite{dwc}. For the 
above left plot, the two macroscopic turn arounds occur in the classical regime. In the right plot, the wiggles on the left occur in the 
non-classical  regime where the magnitude of $\sin(\delta_b b)$ and $\sin(\delta_c c)$ is not close to zero. In the forward evolution, 
the wiggles progressively occur in a more quantum regime.}

\end{figure}

We now investigate whether the energy density is bounded in the effective spacetime description of the $\bar \mu$ quantization. The energy density can be obtained from the Hamiltonian constraint $H_{\bar{\mu}} \approx 0$ as
\be
\rho_{\bar{\mu}}=\frac{1}{8 \pi G \gamma^2 \Delta} \left[2\sin(b\delta_b)\sin(c\delta_c) + \sin^2(b\delta_b)+\frac{\gamma^2 \Delta}{p_c} \right]. \label{rho}
\ee
It is clear that this energy density is bounded for all values of triads and cotriads except when $p_c \rightarrow 0$. Especially, we note that even if the triad $p_b$ is vanishing, the energy density is bounded as far as $p_c$ is nonzero. Since a pancake singularity is attained when $p_c$ remains finite,  we can already conclude that such a singularity is absent in the effective description of the Kantowski-Sachs spacetime for the $\bar \mu$ quantization.\footnote{In contrast, this is not true in the constant $\delta$ and the `improved dynamics inspired' quantizations discussed earlier. For these prescriptions, the expression of energy density contains inverse power of $p_b$ as well as $p_c$ in the expression for energy density. Thus, allowing all kinds of singularities.}

Let us now return to the properties of the energy density in general, and understand its behavior for the generic singularities.   
The energy density in $\bar \mu$ approach will be bounded if $p_c$ does not vanish. In the non-singular evolution, one expects that the dynamics results in a non-zero value of $p_c$. The pertinent question is whether in effective dynamics this happens to be true. Numerical analysis of the Hamilton's equations shows that the answer turns out to be positive. The first evidence of this behavior of $p_c$ was reported in the vacuum Kantowski-Sachs case, where it was found that due to holonomy corrections, $p_c$ (as well as $p_b$) undergo non-singular evolution, and $p_c$ never approaches zero throughout the evolution \cite{bv}. It was found that $p_c$ approaches an asymptotic non-zero value after classical singularity is avoided. Detailed numerical analysis of effective Hamiltonian constraint (\ref{ham}) for different types of matter fields shows that a similar behavior occurs for $p_c$ in general \cite{js2}. An example of this phenomena is shown in Fig. 1,
 where we plot the behavior of $p_c$ versus proper time for the case of massless scalar field in a typical numerical simulation. Giving 
the initial date at $t=0$ we numerically solve the Hamilton's equations for the effective Hamiltonian constraint (\ref{ham}).  During the 
past and future evolution, the physical volume does not go to zero when the classical singularity is approached, but instead bounces. The 
triad $p_c$ never goes to zero in the entire evolution, but asymptotes towards a constant value. These results, and also of Ref. \cite{bv}, 
confirm that dynamically $p_c$ is always bounded away from zero. Hence, we conclude that the energy density (\ref{rho}) is always bounded in 
the loop quantization of the Kantowski-Sachs spacetime.


\section{Conclusions}

Classical Kantowski-Sachs spacetime is singular for generic matter choices, which calls upon a quantum gravitational treatment to see if the singularity persists. A good understanding about the geodesic completeness of a spacetime can be obtained via expansion and shear scalars. Any divergence in these scalars indicates presence of a singularity. Since singularity denotes break down of the theory which is used to describe spacetime, it is hoped that the right theory of quantum gravity will resolve these singularities in general. A quantum theory of spacetime should pass  various consistency tests. If the spatial manifold is non-compact, then the expansion and shear scalars  must be independent of the choice of the fiducial cell. If the singularities are indeed resolved, then the curvature scale associated with singularity resolution should not be arbitrary.
Due to quantization ambiguities, various prescriptions can exist for quantization of a spacetime. Is it possible that a particular prescription is favored over others? This question was earlier posed in the isotropic \cite{cs1} and Bianchi-I spacetime in LQC \cite{cs3}, where it was found that $\bar \mu$ quantization prescription in contrast to other quantization prescriptions leads to generic bounded behavior of expansion and shear scalars, and physical predictions free from the rescaling under fiducial cell. The goal of this analysis was to answer this question in the loop quantization of Kantowski-Sachs spacetime assuming the validity of effective spacetime description for minimally coupled matter.

Previous works on loop quantization of Kantowski-Sachs spacetime have been mostly devoted to study the vacuum case, for which the expansion
scalar has been partially studied earlier \cite{cortez}.  Little details about the physics of singularity resolution for generic matter
were so far available.  Three quantization prescriptions were proposed in the literature. Of these, only one was shown to be preferred in
the sense that the effective Hamiltonian does not depend on the rescalings of the fiducial length. This quantization prescription (denoted
by $\bar \mu$) is the analog of the improved dynamics in isotropic LQC \cite{aps3}. The other two quantization prescriptions, denoted by
$\mu_o$ and $\bar \mu'$ lead to resolution of singularities in the vacuum case, but were known to be problematic under rescalings of the
fiducial cell. Unlike $\bar \mu$ prescription, these  also yield quantum difference equations which are von-Neumann unstable \cite{BCK}. We
obtained the expansion and shear scalars using the effective dynamics in each of these prescriptions and found that except the case of $\bar
\mu$ quantization, in both the other choices these scalars are not necessarily bounded in the effective spacetime. Thus
it is possible that a strong curvature
singularity may not
get resolved for $\mu_o$ and $\bar \mu'$ prescriptions for some choices of matter depending on the initial conditions in effective
dynamics.  Even if the singularities are
resolved, we found that the associated curvature scale is arbitrary. In contrast, the $\bar \mu$ quantization leads to universal bounds on
the expansion and shear scalars which are dictated by the underlying Planckian geometry for Kantowski-Sachs spacetime with matter.
These bounds point towards a generic resolution of sinularities in this prescription. Analysis
of the behavior of energy density in $\bar \mu$ prescription reveals that it is dynamically bounded because $p_c$ is bounded from below. It
turns out that this is a generic feature of all types of perfect fluids, whose details will be reported in a future work \cite{js2}. It is
interesting to note that without solving dynamical equations, it is possible to rule out pancake singularities in the $\bar \mu$
prescription. The bounded behavior of expansion and shear scalars and energy density is a strong indication that curvature singularities may be
generically resolved in the $\bar \mu$ quantization prescription of the Kantowski-Sachs spacetime with matter, as in the case of isotropic
and Bianchi-I model \cite{ps09,sv,ps11}.

To investigate whether there is another quantization prescription which gives a bounded behavior of expansion and shear scalars, we considered a general ansatz of the edge lengths of the holonomies. It turns out that $\bar \mu$ quantization is a unique choice for which the expansion and shear scalars are bounded. For any other prescription, expansion and shear scalars can be unbounded in the effective dynamics. It is remarkable that the demand that these scalars are bounded also chooses the prescription which is free from the rescalings of the fiducial cell. This property is shared by the $\bar \mu$ quantization in the isotropic and Bianchi-I spacetime in LQC \cite{cs1,cs3}. All these similarities between the $\bar \mu$ quantization of the isotropic, Bianchi-I and Kantowski-Sachs spacetimes bring out a harmonious and robust picture of the loop quantization.

Finally, it is important to stress that though this analysis provides further insights on the loop quantization of Kantowski-Sachs spacetime, singling out the $\bar \mu$ prescription on various grounds, more work is needed to rigorously formulate the $\bar \mu$ prescription in the quantum theory. It is known that for the Schwarzschild interior, the $\bar \mu$ quantization results in quantum gravitational effects at the event horizon where the spacetime curvature in the classical theory can be very small \cite{bv}. The existence of these effects is tied to the choice of the coordinates which lead to the classical coordinate singularity at the horizon. Not distinguishing it from the curvature  singularity, quantum geometric effects resulting from the holonomies of the connection components thus become significant at the horizon resolving even the coordinate singularity.
Note that this coordinate artifact does not arise in the Kantowski-Sachs spacetime in presence of matter. These issues will be closely 
examined in the $\bar \mu$ quantization of the Schwarzschild interior \cite{cs4}. Further, it has been reported that the Kantowski-Sachs 
vacuum spacetime in the $\bar \mu$ prescription leads to the Nariai-like spacetime after the bounce in the asymptotic approach 
\cite{bv}.\footnote{It is important to make a distinction here with the classical Nariai spacetime, since in the asympotic approach to  
Nariai-like spacetime, the spacetime is quantum.}\footnote{Before the Narial-like phase is asymptotically approached, the spacetime gives 
birth to baby blackhole spacetimes \cite{dwc}.} It turns out that this feature is more general, which reveals some subtle properties of the 
effective spacetime in LQC \cite{js3}. A deeper understanding of these issues is required to gain further insights on the details of the 
physics of singularity resolution in the Kantowski-Sachs spacetime in LQC.

\begin{acknowledgments}
We thank Alejandro Corichi, Naresh Dadhich, Brajesh Gupt, Jorge Pullin and  Edward Wilson-Ewing for  useful discussions. We thank an anonymous referee for useful comments on the manuscript. This work is
supported by a grant from John Templeton Foundation and by NSF grants PHYS1068743 and PHYS1404240. The opinions expressed
in this publication are those of authors and do not necessarily reflect the views of John Templeton Foundation.
\end{acknowledgments}

\end{document}